\documentstyle[12pt,epsfig]{article}
\begin{document}

\newcommand{\be}{\begin{equation}}
\newcommand{\ee}{\end{equation}}
\begin{center}

{\bf High energy Cherenkov gluons at RHIC and LHC}\\

\vspace{2mm}

I.M. Dremin\footnote{Email: dremin@lpi.ru}, 
L.I. Sarycheva\footnote{Email: lis@alex.sinp.msu.ru}, 
K.Yu. Teplov\footnote{Email: teplov@lav01.sinp.msu.ru}\\

\vspace{2mm}

$^1${\it Lebedev Physical Institute, Moscow}\\
$^2$$^,$$^3$ {\it D.V. Skobeltsyn Institute of Nuclear Physics,\\
                 M.V. Lomonosov Moscow State University, Moscow}\\

\end{center}

\begin{abstract}
The collective effect of emission by the forward moving partons of high 
energy Cherenkov gluons in nucleus-nucleus collisions at RHIC and LHC 
energies is considered. It can reveal itself as peaks in the pseudorapidity 
distribution of jets at midrapidities or as a ring-like structure of
individual events in event-by-event analysis.
The pseudorapidity distribution of centers of dense isolated groups
of particles in HIJING model is determined. It can be considered as the
background for Cherenkov gluons. If peaks above this background were found
in experiment, they would indicate new collective effects.
\end{abstract}

The search for collective effects in hadronic and nuclear high-energy
reactions has always been one of the mainstreams of experimental and 
theoretical investigations. Among them, Cherenkov gluons \cite{d1, d2} 
and Mach waves \cite{glas} were discussed long ago. Cherenkov
gluons are the intuitive analogue to Cherenkov photons if the
electron beam is replaced by the bunch of partons (quarks and gluons)
traversing the nuclear medium. In their turn, Mach waves could appear
in this medium if the parton (jet) speed exceeds the speed of sound.
Recently, the interest to these effects was revived 
\cite{dim,mwa,kmwa,shur,stoc,mrup} in connection with RHIC data 
\cite{wang} as reviewed in \cite{dim}. The common feature of these 
collective coherent processes is particle production concentrated 
on a cone with the polar angle $\theta $ defined by the condition
\begin{equation}
\cos \theta = \frac{c_w}{v},     \label{cos}
\end{equation}
if the infinite medium at rest is considered and the direction of the parton 
motion is chosen as the cone axis. Here $v$ is the velocity of the 
parton producing these effects, $c_w$ is the phase velocity of gluons,
or sound velocity in the medium. 

The main experimental signature of both effects would be two peaks in 
the pseudorapidity distribution of particles produced in high energy 
nuclear collisions which are positioned in accordance with Eq. (\ref{cos}).
The most visual image of these effects is the ring-like structure
of events in the plane perpendicular to the direction of propagation
of the body initiating them.

At high energies of initial partons (jets) $v\approx c$. The velocity $c_w$ can
range from quite low values to a value slightly below $c$ according to present 
estimates for  both effects (see \cite{dim}). The lowest values of $c_w$ are 
obtained for rarefied media and low energy gluons, while larger $c_w$ 
correspond to strong shock waves and high energy gluons. 

For gluons, $c_w=c/n$ where $n$ is their nuclear index of refraction in 
a nuclear matter through which they move. The necessary condition for this 
effect is that the real part of the index of refraction be larger than 1.
This index was estimated from experimental data on hadronic reactions 
\cite{d2, dim} with assumption that gluons as carriers of strong forces
should possess the features common to hadronic reactions. Its value is 
proportional to the real part of the forward scattering amplitude, and we 
know from experiment (and from its Breit-Wigner and dispersion relations
theoretical description) that for any hadronic process it becomes 
positive in presence of any resonance and at very high energies.
Thus, the necessary condition is satisfied in these cases and one can wait
for observable effects with low energy and high energy gluons.

For low energy gluons which can generate hadronic resonances, the real 
part of the nuclear index of refraction can be written \cite{dim} as
\begin{equation}
{\rm Re} n^r=1+\Delta n_R^r=1+\frac {3m_{\pi }^3}{2\omega _r^2\Gamma}. 
\label{del}
\end{equation}
Here $\omega _r$ is the energy required to produce a resonance. It can be of 
the order of the pion mass $m_{\pi }$. Since the widths of known resonances $\Gamma$ are of the 
order of hundred MeV, $\Delta n_R^r$ can be of the 
order of 1. Therefore, according to (\ref{cos}), the angle of particles 
emission is rather large in the target rest system. The effect can be observed
at RHIC and LHC if initial partons (jets) move at a large angle with 
respect to the collision axis. In such a way one can try to interpret 
the recently observed at RHIC \cite{wang} effect with two peaks in angular 
distribution about the direction of propagation of the companion jet
created in the direction perpendicular to the collision axis.
The peak position showed that $c_w=0.33c$. Thus, it could be the emission 
of low energy Cherenkov gluons with nuclear index of refraction equal to 3.
In this case the resonance production should be enlarged in this angular 
region. It can result in different ratio of pions to protons compared to that
outside this region.
In \cite{shur} it was interpreted as Mach waves with $c_w=0.33c$. 
However, the special trigger is needed to observe this effect as it was done
in \cite{wang}. Moreover, the production of the trigger and companion jets at 
90$^\circ$ is rather rare process which requires the high statistics experiment.
This effect is unobservable at RHIC and LHC for the forward moving partons
because in this case the large emission angles in the target rest system are 
trasformed to angles extremely close to $\pi $ in RHIC and LHC systems.
However, it could be observable for forward moving partons in fixed target 
experiments as peaks at about 70$^\circ$ but, strangely enough, no such observations have
yet been presented.

The impinging nuclei can be considered as bunches of the forward/back\-ward moving high 
energy partons passing through each other. Beside unobservable at RHIC and LHC
(as explained above) low energy gluons, each initial forward moving parton can emit high energy 
Cherenkov gluons when traversing the target nucleus (as well as target partons can do the same in 
the opposite hemisphere). The real part of the nuclear index of refraction
has been estimated \cite{d2} using the formula 
\begin{equation}
{\rm Re} n^h(\omega )=1+\Delta n_R^h(\omega )=1+\frac {3m_{\pi }^3}{8\pi \omega}
\sigma (\omega )\rho (\omega ),     \label{ren}
\end{equation}
where $\rho (\omega )={\rm Re} F/{\rm Im} F$, $F(\omega)$ and $\sigma (\omega)$ being
the hadronic forward scattering amplitude and the cross section. It becomes positive
above some threshold, increases and then decreases at high energies $\omega $ 
so that
\begin{equation}
\Delta n_R^h(\omega )\approx \frac {a}{\omega}~,     \label{dhig}
\end{equation}
where $a\approx 2\cdot 10^{-3}$ GeV  if $\rho \approx 0.1$ as it follows from 
experiment, and assumed to remain constant at higher energies. The index $h$ refers 
to high energy gluons. Therefore, according to (\ref{cos}), the angle of 
particles emission is quite small in the target rest system but much larger
than bremsstrahlung angles. If transformed to RHIC or LHC systems, these 
angles can become large (somewhere in the midrapidity region). The effect can 
be observed if initial it partons (jets) move (almost) along the collision 
axis. There are numerous experimental indications in favor of this effect 
(see review in \cite{dim}). The first one of them was presented in \cite{addk}.
Most results are, however, either for individual cosmic rays events or for 
special samples of events at accelerator energies.

Here we should mention that the finite length of nuclear targets can change 
somewhat the estimate (\ref{cos}): enlarge the transverse 
momenta of particles in Cherenkov jets and influence the difference
between processes with different colliding nuclei \cite{d2, dim}.

In what follows we discuss high energy Cherenkov gluons at RHIC and LHC 
energies produced by forward moving partons. The important problem of 
experimental search for this effect is the shape of the
background due to ``ordinary'' processes. Its influence should be minimized.
For doing this we propose to use the distinctive feature of production of
high energy Cherenkov gluons. Namely, such gluon should produce a jet of 
particles which can be distinguished as a high density isolated group
of particles. Therefore, the distributions of groups (jets) of particles 
should be considered rather than inclusive particle distributions.
Separating such groups from experimental data one would 
increase the relative contribution of jets produced by Cherenkov gluons.
By such selection we exclude weakly correlated particles. Statistical 
fluctuations and hard QCD-jets are still accounted for but their relative 
probability is reduced and pseudorapidity distribution must be rather smooth.
Therefore the role of background in the distribution of the centers of 
such groups becomes lower compared to the overall pseudorapidity 
distribution. Peaks corresponding to Cherenkov gluons should be more 
pronounced. 

To estimate the background we have used the HIJING model for central
collisions ($b=0$) for Au--Au collisions at RHIC energy
$\sqrt s$=200A GeV and for Pb--Pb collisions at LHC energy $\sqrt s$=5500A GeV.
3500 events were generated in each case. We have chosen the central
collisions because the number of forward moving participants and, 
correspondingly, their role is larger in central collisions of heavy nuclei. 
Moreover, the Cherenkov radiation intensity is proportional to the length of the parton 
path in the medium. It is also larger in central collisions. 

\begin{figure} [hbtp]
\begin{center}
\vspace{-50mm}
{\epsfig{file=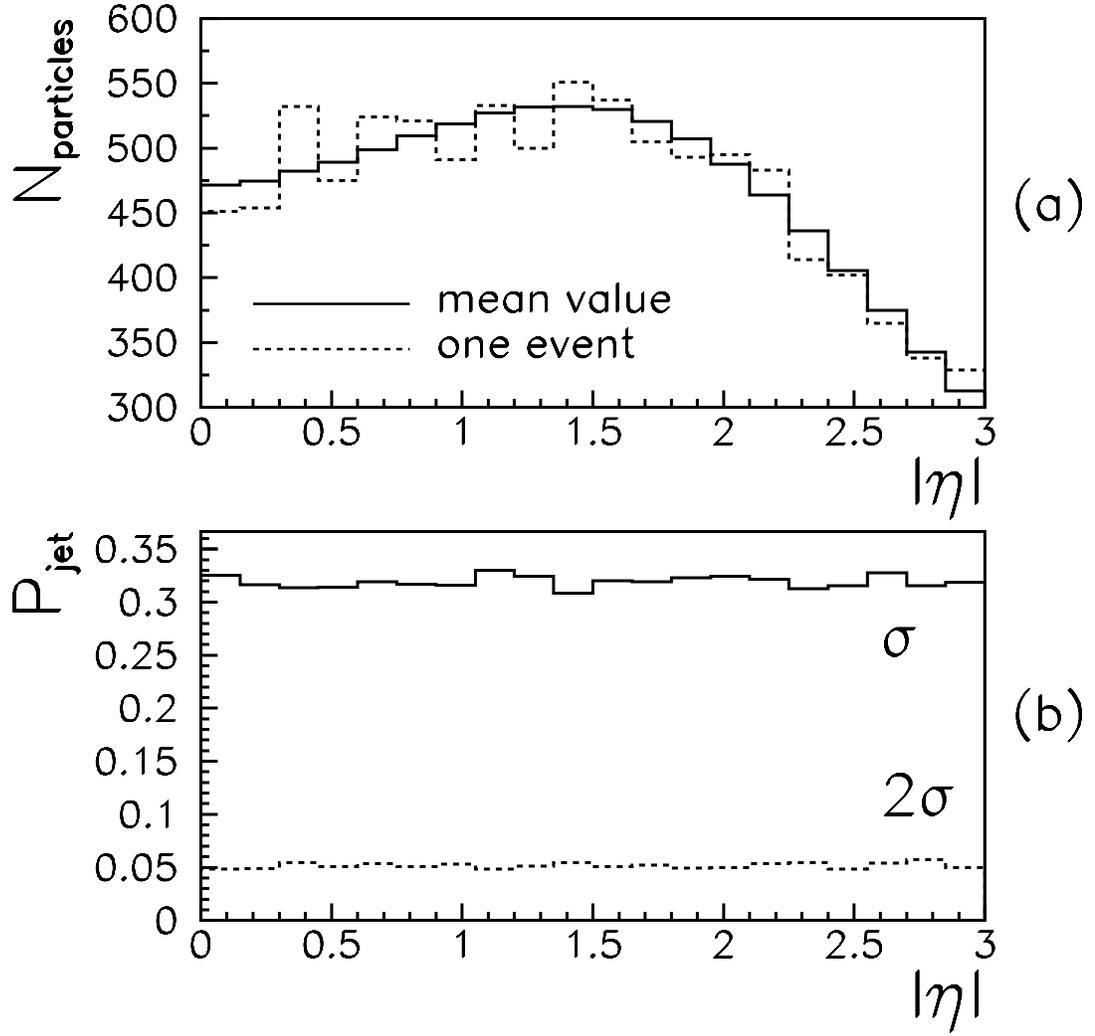,width=15cm,height=15cm}}
\end{center}
\caption{(a) The pseudorapidity distribution in one of HIJING events 
(dashed histogram) for central Au--Au collision at $\sqrt s$=200A GeV is plotted 
over the inclusive HIJING distribution (solid histogram), 
 $N_{particles}$ --- number of particles. Peaks above the 
inclusive plot are clearly seen.
(b) The pseudorapidity distribution of the centers of dense
isolated groups of particles similar to those shown in Fig. 1a and 
exceeding the inclusive plot by two and one standard 
deviations $\sigma$, $P_{jet}$ --- probability to find peak
above mean + $\sigma$ ($2\sigma$).
This is the smooth background for further searches
of collective effects.} 
\label{fig:pic1}
\end{figure}

\begin{figure} [hbtp]
\begin{center}
\vspace{-50mm}
{\epsfig{file=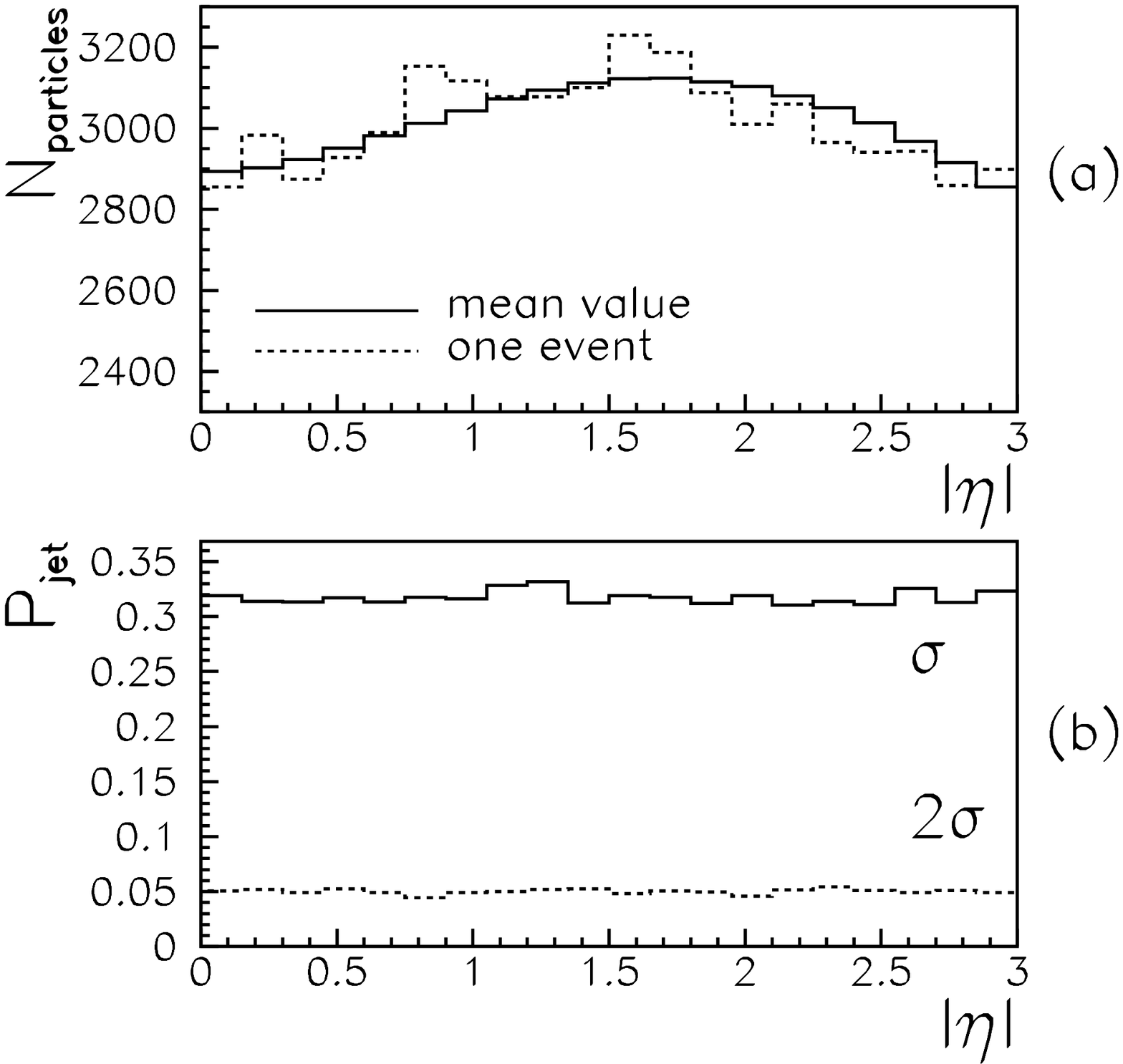,width=15cm,height=15cm}}
\end{center}
\caption{
(a) The pseudorapidity distribution in one of HIJING events 
(dashed histogram) for central Pb--Pb collision at $\sqrt s$=5500A GeV is plotted 
over the inclusive HIJING distribution (solid histogram), 
 $N_{particles}$ --- number of particles. Peaks above the 
inclusive plot are clearly seen. (b) The pseudorapidity distribution of the centers of dense
isolated groups of particles similar to those shown in Fig. 2a and 
exceeding the inclusive plot by two and one standard 
deviations $\sigma$, $P_{jet}$ --- probability to find peak
above mean + $\sigma$ ($2\sigma$). This is the smooth background for further searches
of collective effects.} 
\label{fig:pic2}
\end{figure}

Then the peakes in individual HIJING events exceeding the regular distribution
by more than one and two standard deviations have been separated. They
can appear either as purely statistical fluctuations or as hard QCD-jets.
Figs \ref{fig:pic1}a and \ref{fig:pic2}a show the examples of such events (for RHIC and 
LHC energies, correspondingly) plotted over the smooth inclusive
pseudorapidity distributions. 

Peaks exceeding the distributions are clearly seen. All simulated events
have been plotted in such a way and centers of peaks defined.
Finally, the distribution of the centers of these peaks was plotted.
Figs \ref{fig:pic1}b and \ref{fig:pic2}b show these distributions for peaks
exceeding the inclusive background
at RHIC and LHC energies by two or one standard deviations. It is seen that
these distributions are flat with extremely small irregularities. This
agrees with our expectations that statistical fluctuations and QCD jets do
not have any preferred emission angle and should be randomly dispersed over the 
inclusive particle distribution. They can be considered as a background 
for experimental search for Cherenkov gluons which do have such 
preferred angle. High energy Cherenkov jets should have quite narrow angular 
spread. If their angular width corresponds to a single bin in Figs. 1 and 2,
then they would produce peaks twice exceeding this background even when 
their cross section is only 5 per cent of the cross section in the
considered interval of pseudorapidities.
If experimental data on group centers distribution show 
some peaks at definite pseudorapidity values over this background,
this can be indicative of new collective effect, not considered
in HIJING. These findings may be added to the experimental evidence in favor 
of such effect collected before (they are reviewed in \cite{dim}).

It is easy to check from Figs that the levels of the background for 1$\sigma $
and 2$\sigma $ fluctuations correspond to the traditional statistical
estimates of about 30$\%$ and 5$\%$. However, in principle, these levels 
could be counted, e.g., from the inclusive pseudorapidity distribution, so
the background would have certain distinct shape. It should be stressed once 
again that the main result of the paper consists in the constancy of this 
background. The presence of the traditional QCD jets in HIJING does not change 
it. This simplifies experimental task of search for deviations from the flat 
distribution.

To conclude, the pseudorapidity distributions of the centers of dense 
isolated groups of particles (jets) exceeding in individual events the 
inclusive distribution are plotted for events generated according to HIJING 
model at RHIC and LHC energies. They are very flat and provide the background 
for further searches for such collective effects as Cherenkov gluons and
Mach waves. If the peaks in the pseudorapidity plot of the centers of 
separated groups are found in experiment and fit the condition 
(\ref{cos}), then it will testify in favor of a Cherenkov gluons hypothesis.
The positions of the peaks reveal such property of hadronic matter
as its nuclear index of refraction and can be valuable for understanding
the equation of state of the nuclear medium. \\

{\bf Acknowledgements}\\

This work has been supported in part by the RFBR grants 03-02-16134, 
04-02-16445-a, 04-02-16333, NSH-1936.2003.2. Authors are grateful to 
A. Demianov for useful discussions.\\

\end{document}